\documentclass[a4paper,11pt]{article}
\pdfoutput=1 

\usepackage{jinstpub_modified} 
                     
\usepackage[LGRgreek]{mathastext}

\title{\boldmath R\&D studies on eco-friendly gas mixtures for the ALICE Muon Identifier }

\author[a, 1]{Antonio Bianchi,\note{Corresponding author.}}

\affiliation[a]{INFN and University of Torino, Via Pietro Giuria 1, 10125, Torino, Italy}

\emailAdd{antonio.bianchi@unito.it}

\abstract{Resistive Plate Chambers (RPCs), used for the Muon Spectrometer of the ALICE experiment at CERN LHC, are currently operated in maxi-avalanche mode with a low threshold value and without amplification in the front-end electronics. RPC detectors have shown a good operation stability with the current gas mixture during the entire Run 1 (2010$-$2013) and the ongoing Run 2 (2015$-$2018) at the LHC.

The gas mixture is made up of $C_{2}H_{2}F_{4}$, $SF_{6}$ and $iC_{4}H_{10}$. Since the first two gases have high Global Warming Potentials (GWPs), there is the risk that they will be phased out of production in the next years, due to the recent restrictions and regulations of the European Union; meanwhile their cost is progressively increasing. Therefore, finding a new eco-friendly gas mixture has become extremely important in order to reduce the emissions of greenhouse gases. In addition, the present $iC_{4}H_{10}$ contribution makes the current gas mixture flammable. Non-flammable components, or at least in non-flammable concentrations, would be advisable to make the operation of detectors simpler and safer.

In order to identify a gas mixture with the improved characteristics and suited to cope with the requirements of the ALICE Muon Identifier in the forthcoming High-Luminosity runs, a dedicated experimental set-up has been used to carry out R\&D studies on promising gas mixtures with small-size RPCs.

Hydrofluoroolefins ($HFOs$) are appropriate candidates to replace the $C_{2}H_{2}F_{4}$ thanks to their very low GWPs, especially $HFO1234ze$ which is not flammable at room temperature. Several tests on $HFO$-based mixtures with addition of various gases are ongoing and encouraging results have already been obtained. Furthermore, the use of $CO_{2}$ as a quencher has been studied as it might represent a valid solution to avoid flammability of the mixture. Finally, medium-term stability of detectors exposed to the cosmic-ray flux will be shown in this paper.}


\collaboration[c]{on behalf of the ALICE collaboration}

\proceeding{14$^{\text{th}}$ Workshop on Resistive Plate Chambers and Related Detectors\\
  when 19-23 February 2018\\
  where Puerto Vallarta, Jalisco State, Mexico}

\begin{document}
\maketitle
\flushbottom

\section{Introduction}
\label{sec:intro}
ALICE (A Large Ion Collider Experiment~\cite{a}) is a general-purpose heavy-ion experiment at the Large Hadron Collider (LHC). The experiment was designed to study ultra-relativistic heavy ion collisions in order to detect and characterize a new state of matter: the Quark$-$Gluon Plasma (QGP).

The measurement of heavy-flavored hadrons (containing a quark charm or beauty) and quarkonia (bound state $c\overline{c}$ and $b\overline{b}$) is fundamental in this study, since the probes provide information on the interaction of heavy quarks with the medium and on the status of confinement in the QGP.
Heavy flavors and quarkonia are measured in their muonic decay channels with the ALICE Muon Spectrometer, which consists of a complex arrangement of absorbers, a large dipole magnet, and twelve stations of tracking and triggering chambers. 

The Muon Trigger system of the experiment is currently made up of 72 bakelite Resistive Plate Chambers (RPCs~\cite{b}) with a 2 mm single-gap. The RPCs are arranged in four detection planes and equipped with front-end electronics~\cite{g} with a low threshold value (7 mV) and without amplification. The RPCs are currently operated in maxi-avalanche mode with a gas mixture of 89.7\% $C_{2}H_{2}F_{4}$, 10\% $iC_{4}H_{10}$ and 0.3\% $SF_{6}$, showing a good performance in the LHC environment~\cite{c}. Water vapor is added to the gas mixture in order to maintain the relative humidity equal to about 45\% and avoid any changes in the resistivity of the bakelite RPC electrodes.

The current gas mixture is not environmental friendly because of the presence of greenhouse gases; they are classified according to their Global Warming Potential (GWP), which is a relative measurement of how much heat is trapped into the atmosphere with respect to the same mass of $CO_{2}$. Indeed the $C_{2}H_{2}F_{4}$ has a GWP equal to 1430, whereas the GWP of $SF_{6}$ is 22200. Since January 2015, a new European Union (EU) regulation~\cite{d} has limited the total emissions of fluorinated gases or even banned their use when less harmful alternative are already available. The EU regulation might have a significant effect on future availability of these gases and their prices could rise. In addition, the use of $10\% \ iC_{4}H_{10}$ makes the current ALICE mixture flammable.
The gas volume for the Muon Trigger is quite small (about $0.3 \  m^{3}$), nevertheless an alternative gas mixture is welcome.

R\&D studies on non-flammable mixtures without greenhouse gases have been performed for possible future upgrades in view of the forthcoming ALICE runs at the LHC. The ongoing experimental approaches are focused on the reduction of the total GWP and the removal of flammable components in the current gas mixture.

\section{RPC operation with non-flammable and environmental friendly gases
}
The GWP of the current mixture is equal to 1350: 95\% of this is due to the presence of $C_{2}H_{2}F_{4}$. The hydrofluoroolefins may be appropriate candidates to replace $C_{2}H_{2}F_{4}$, thanks to their very low GWPs. Among the available hydrofluorooolefins the tetrafluoropropane ($C_{3}H_{2}F_{4}$), which can be found with the trade-name $HFO1234ze$ ($HFOze$), may represent a valid alternative to the $C_{2}H_{2}F_{4}$: its chemical structure suggests a good capability to absorb photons, due to the presence of a double bond in the chain of carbon atoms, and its GWP is lower than 1. In addition, $HFOze$ is not flammable at room temperature and recent studies~\cite{e} suggest a strong electron attachment.

The direct replacement of $HFOze$ with $C_{2}H_{2}F_{4}$ would need an increase of the electric field, leading to operating voltages larger than 14 kV, which is not advisable for the electrical stability of our system. Therefore several tests with the addition of different gases have been performed in order to operate at lower values of the electric field and identify some promising $HFO$-based mixtures for the requirements of the Muon Identifier in the forthcoming High-Luminosity runs. The addition of $CO_{2}$ is the most promising so far, therefore only results on $HFO$-based mixtures with $CO_{2}$ will be shown in this work.

\subsection{Characterization of eco-friendly gas mixtures for RPC operation}
\subsubsection{Experimental set-up}
\label{sec:exp}
\begin{figure}[htbp]
\centering
\includegraphics[width=.7\textwidth,clip]{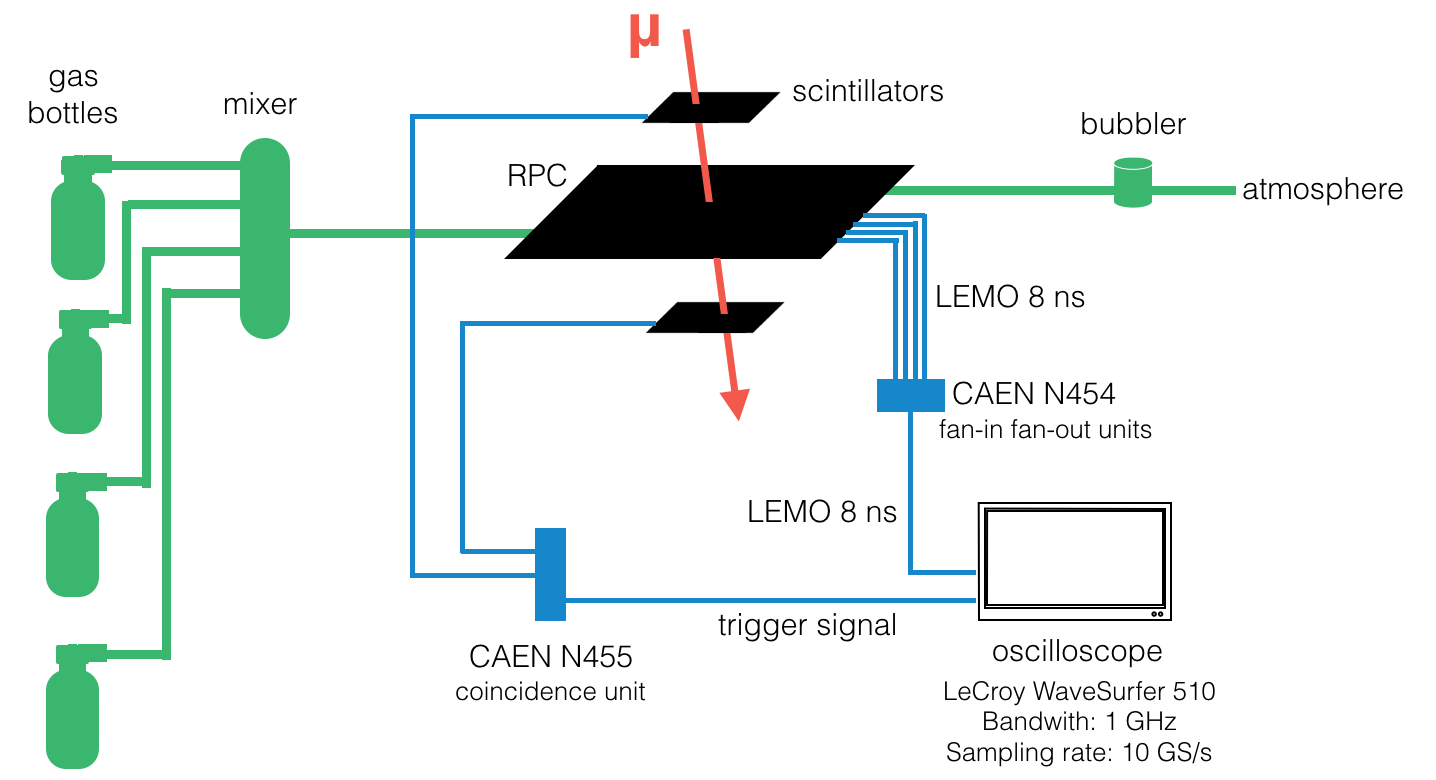}
\caption{\label{fig:1} Experimental set-up for R\&D studies on eco-friendly gas mixtures.}
\end{figure}

A dedicated experimental set-up, shown in Fig.~\ref{fig:1}, has been used to carry out R\&D studies on various gas mixtures flushing two small-size ($50 \cdot 50 \ cm^{2}$, $2 \ mm$ gas gap) RPCs, exposed to the cosmic-ray flux. The muon trigger is provided by the coincidence of four scintillators, coupled with photomultipliers. High voltages (HV) are applied to the detectors with temperature and pressure correction. The gas mixture for the performance measurements is dry and, consequently, the resistivity of the bakelite electrodes is kept optimal by flowing the RPCs with a wet mixture every 10/15 days. Signals, induced on the about 2 cm wide read-out strips, are acquired by an oscilloscope (LeCroy WaveSurfer 510) with a bandwidth equal to 1 GHz and sampling rate at 10 GS/s. The signal waveforms are divided into three different time windows for the analysis: in the first one with 100 ns width the voltage baseline is determined, in the second one (100 ns) the amplitude and the charge of the signals are measured and finally in the last one (200 ns) the possible presence of after-pulses is checked.

The detector is considered as efficient to the passage of a particle if the corresponding signal has an amplitude larger than 3 mV and a charge larger than 0.3 pC. If the measured charge is larger than 3 pC, the signal is considered as a streamer in this work.

\subsubsection{Characterization of $HFO$-based mixtures with addition of $CO_{2}$
}
\begin{figure}[htbp]
\centering
\includegraphics[width=0.9\textwidth,clip]{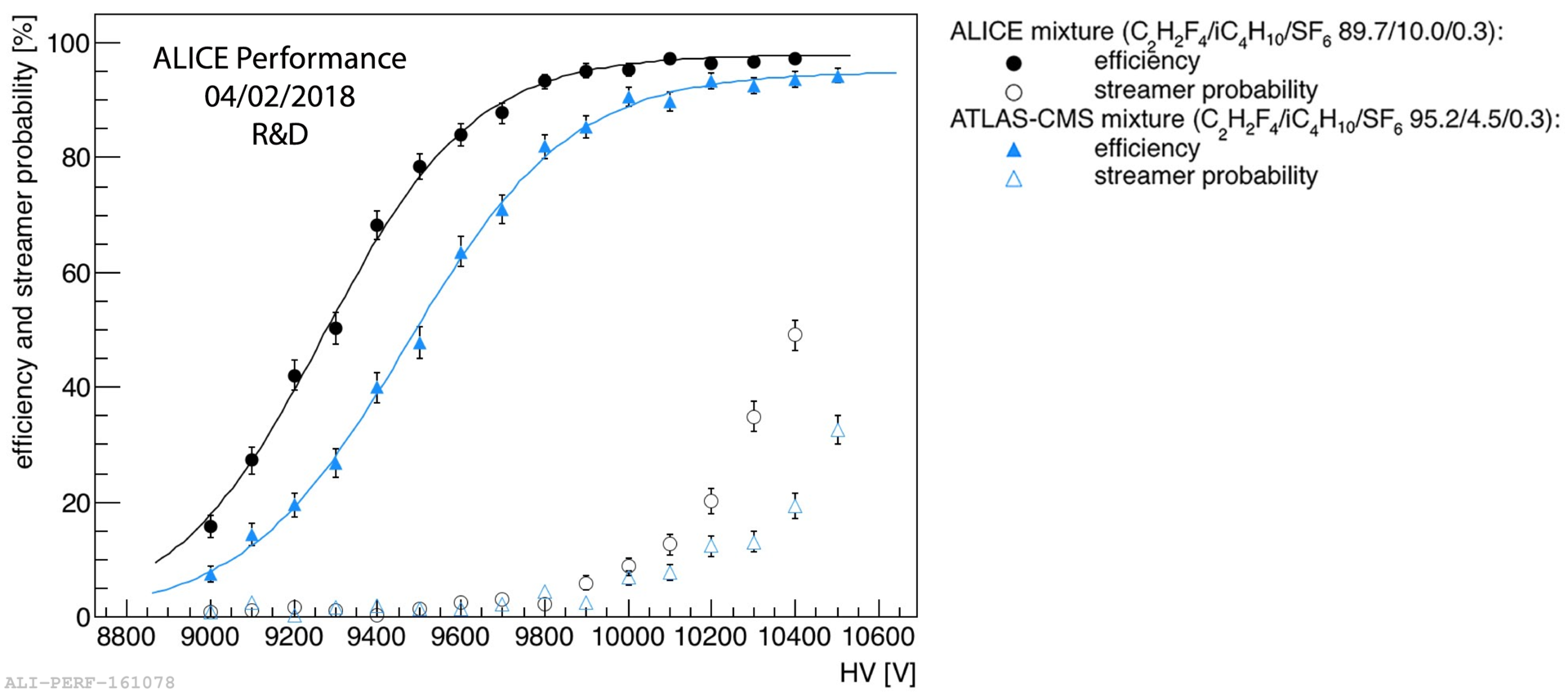}
\caption{\label{fig:2} Efficiency curve and streamer probability for current and ATLAS$-$CMS gas mixtures.}
\end{figure}

Before starting studies on detector performance with $HFO$-based mixture, RPCs have been characterized with the most common gas mixtures at the LHC for this type of detector. Fig.~\ref{fig:2} shows the efficiency curve and the streamer probability with the current ALICE and ATLAS$-$CMS gas mixtures.

\begin{figure}[htbp]
\centering
\includegraphics[width=0.9\textwidth,clip]{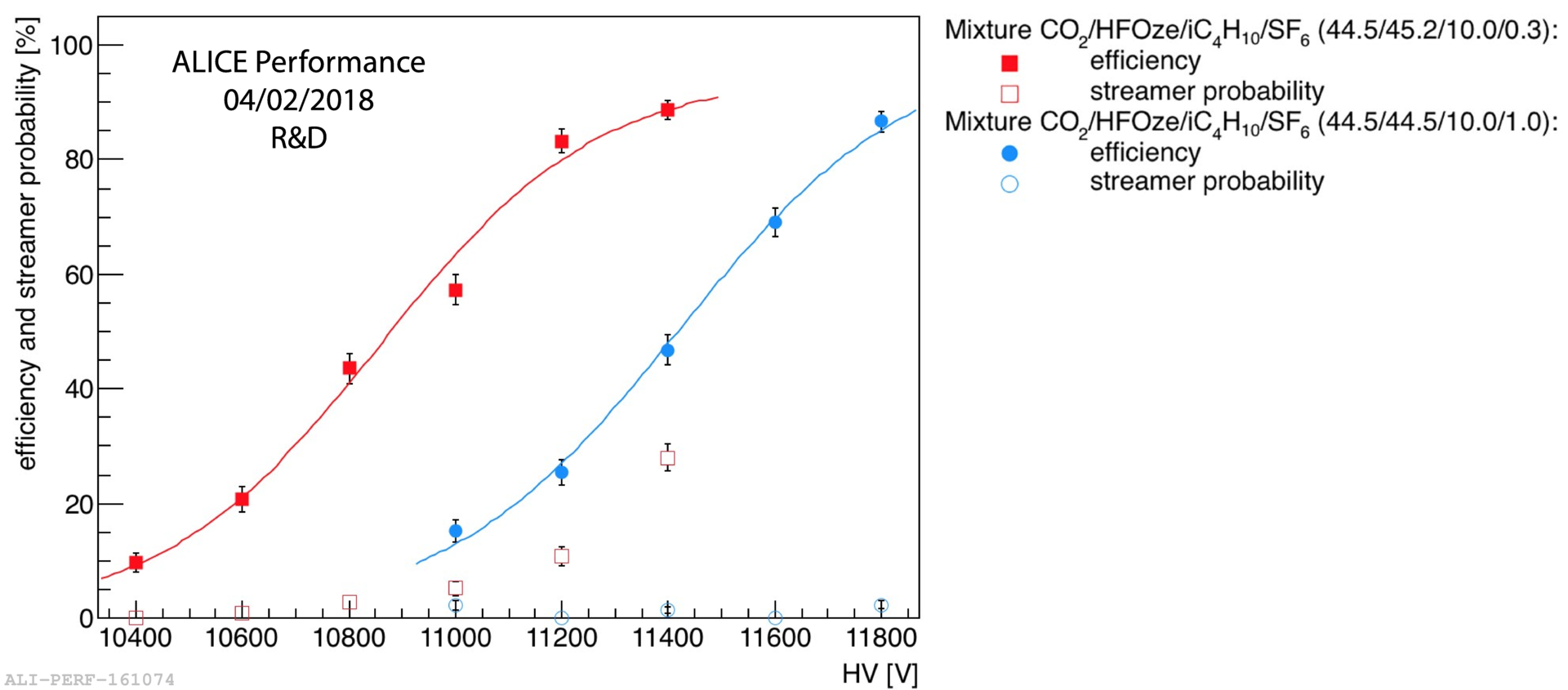}
\caption{\label{fig:3} Efficiency and streamer probability for HFO-based gas mixtures with 0.3\% and 1.0\% $SF_{6}$.}
\end{figure}

In the next step, the fraction of $C_{2}H_{2}F_{4}$ (89.7\%) that is currently present in the gas mixture was replaced with 44.5\% $CO_{2}$ and 45.2\% $HFOze$. The efficiency curve is shifted to higher values by 1500 V in comparison to the detector efficiency with the present gas mixture and the streamer probability is increased, as shown in Fig.~\ref{fig:3}. The addition of $SF_{6}$ reduces the streamer probability: in particular, a value lower than 5\% is observed at an efficiency equal to 90\% when the mixture includes 1\% $SF_{6}$, even if the working point is shifted by another 500 V.

\begin{figure}[htbp]
\centering
\includegraphics[width=0.9\textwidth,clip]{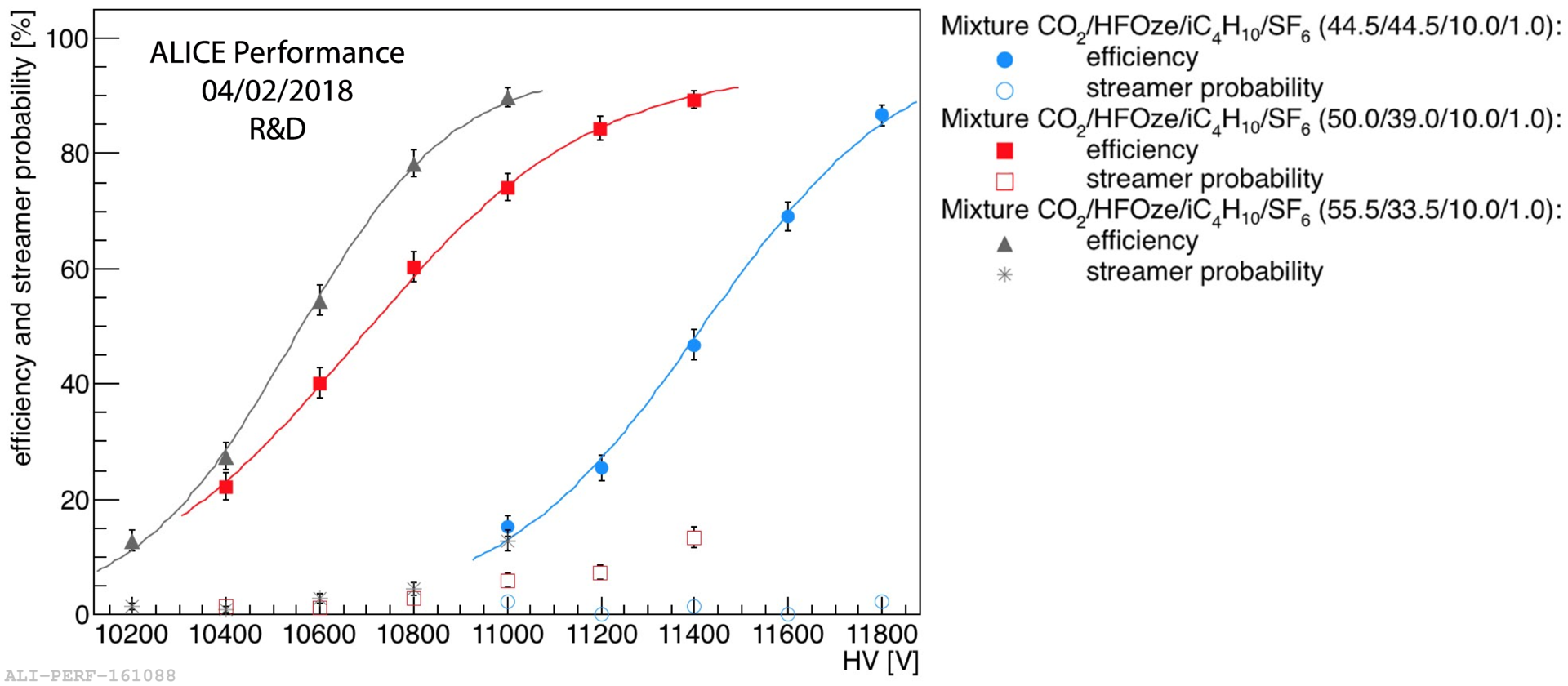}
\caption{\label{fig:4} Efficiency and streamer probability for gas mixtures with different ratios between $CO_{2}$ and $HFOze$.}
\end{figure}

The ratio between $CO_{2}$ and $HFOze$ proves to be fundamental in reducing the operating voltage. Fig.~\ref{fig:4} shows the effect of partially replacing $HFOze$ by $CO_{2}$: the working point is progressively shifted to lower values, even if the streamer probability becomes higher. Moreover, an instability of the electric current at the working voltage was observed when a gas mixture with 55.5\% $CO_{2}$ is used in the detector. Further investigations are needed to understand the reasons for such a behavior.

The results obtained with the gas mixtures with 44.5\% and 50\% $CO_{2}$, shown respectively in blue and red in Fig.~\ref{fig:4} are promising and their total GWPs (equal to 223) are almost six times lower than the presently used mixture (equal to 1350). On the other hand, the working point remains high and the contamination of streamers at the operating voltage may be excessive. However the new front-end electronics~\cite{f} for the forthcoming upgrade in view of Run 3 provides an amplification stage which allows one to achieve the maximum detector efficiency at lower values of electric field, in which streamers have smaller chances to develop.

\subsection{Characterization of gas mixtures without flammable components}
The current mixture has a high concentration of $iC_{4}H_{10}$, which makes it flammable. Non-flammable components would be advisable to make the operation of detectors simpler and safer. R\&D studies to avoid the use of flammable components in the gas mixtures are ongoing and especially the replacement of $iC_{4}H_{10}$ with $CO_{2}$ in $C_{2}H_{2}F_{4}$-based mixtures has been investigated.
\\
The experimental set-up is the same as the one presented in the  section~\ref{sec:exp} with only two changes:
\begin{itemize}
\item the data acquisition has been performed with the standard ALICE front-end electronics, which has a voltage threshold equal to 7 mV without amplification;
\item the gas mixture bubbles in water at 10 $^\circ$C before flowing into the detector, whereas the only replacement in the composition is the use of $CO_{2}$ instead of $iC_{4}H_{10}$.
\end{itemize}

\begin{figure}[htbp]
\centering
\includegraphics[width=1\textwidth,clip]{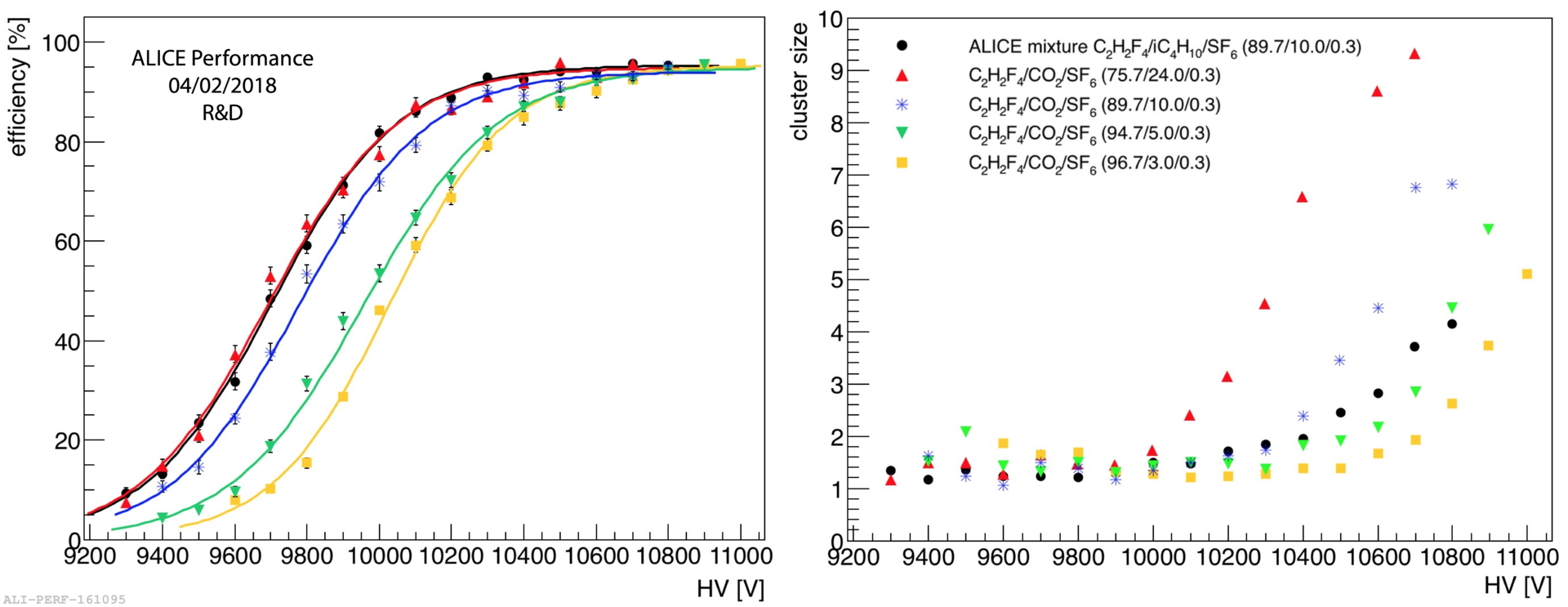}
\caption{\label{fig:5} Efficiency (left) and cluster size (right) for different concentrations of $CO_{2}$ in $C_{2}H_{2}F_{4}$-based mixtures. The error bars for the values of cluster size are hidden by the markers.}
\end{figure}

After the characterization of the RPCs with the current gas mixture, several mixtures have been tested with different ratio of $CO_{2}$ and $C_{2}H_{2}F_{4}$. The resulting efficiency curves and cluster sizes are shown in the left and right side of Fig.~\ref{fig:5}, respectively.

\begin{figure}[htbp]
\centering
\includegraphics[width=1\textwidth,clip]{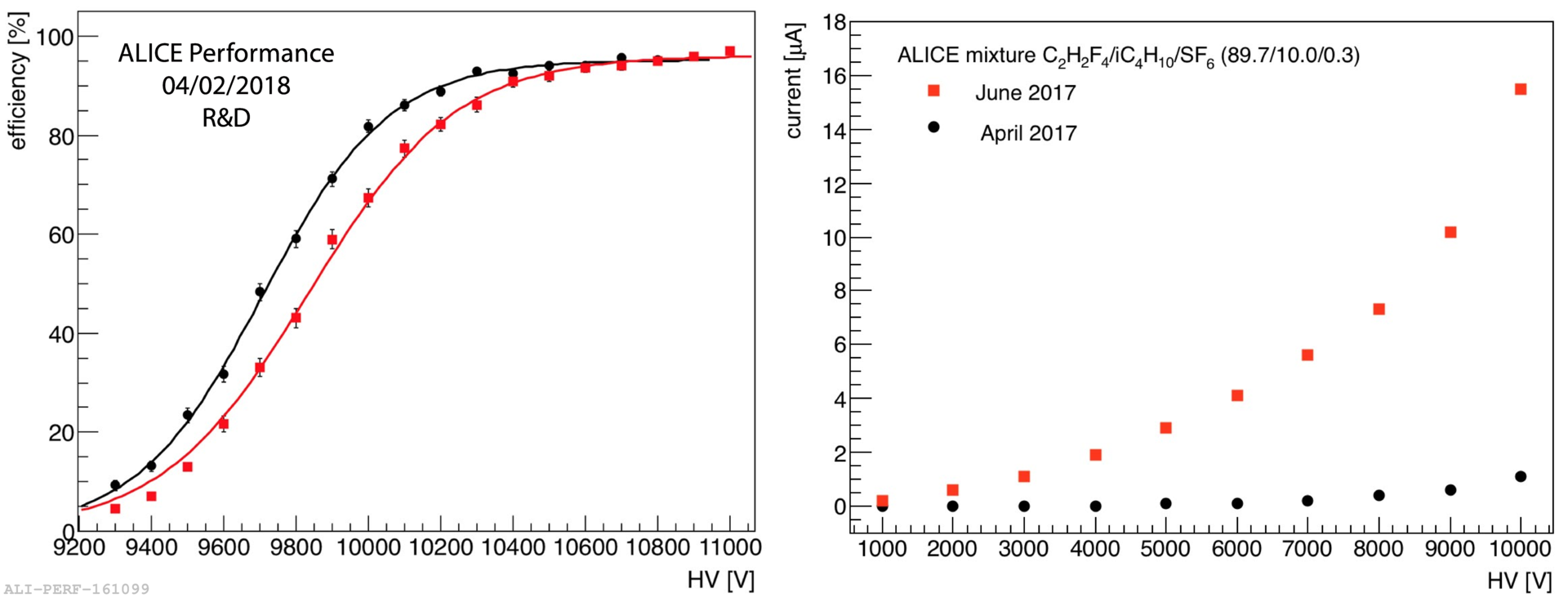}
\caption{\label{fig:6} Efficiency (left) and current absorbed by the RPC detector (right) as a function of the applied voltage measured in April and June 2017. The error bars for the measured currents are hidden by the markers.}
\end{figure}

$C_{2}H_{2}F_{4}$-based mixtures with an addition of $CO_{2}$ instead of $iC_{4}H_{10}$ result in a cluster size at working point generally larger than the one obtained with the present mixture. The cluster size as well as the number of after-pulses observed with the oscilloscope progressively increase with the $CO_{2}$ content. In addition, some stability problems of RPCs have been observed after about two months of the operation.
Figure~\ref{fig:6} shows the shift of the efficiency curve and the huge increase in current by a factor of about 15 in a short period of operation. This issue on the operation stability is under further investigation; possible reasons might be the gas quality, the environmental conditions, the humidity in the mixture, chemical reactions between $CO_{2}$ and the inner surface of the detectors or among the components of the gas mixture, etc.

\section{Conclusions}
R\&D studies on gas mixtures for RPC detectors are ongoing in order to identify and characterize a non-flammable gas mixture without greenhouse gases or at least with a low GWP.
$HFOze$ is a possible candidate to substitute the $C_{2}H_{2}F_{4}$, thanks to its very low GWP. In addition the partial or total replacement of $iC_{4}H_{10}$ with a non-flammable gas would be advisable for security reasons.

Several tests on $HFO$-based mixtures with the addition of various gases are ongoing in order to operate at lower values of electric field. Encouraging results with the addition of $CO_{2}$ have been obtained, but cluster size and time resolution of the detectors are not still measured with the upgraded electronics for the forthcoming data taking period (Run 3). Furthermore long time stability tests are required to check the rate capability, the effects of background radiation, the performance and possible aging effects.

Finally R\&D studies to select only non-flammable components in the gas mixtures are ongoing. The replacement of $iC_{4}H_{10}$ with $CO_{2}$ in $C_{2}H_{2}F_{4}$-based mixtures is not promising because $CO_{2}$ leads to large cluster sizes at the working point. In addition some issues on the operation stability of RPCs with $CO_{2}$ have been observed after two months of operation with cosmic-ray flux. Therefore further tests are fundamental to check the compatibility of $CO_{2}$ in the gas mixture for RPCs.


\end{document}